# Phase Separation and Superparamagnetism in the Martensitic Phase of $Ni_{50-x}Co_xMn_{40}Sn_{10}$


S. Yuan,[1] P.L. Kuhns,[1] A.P. Reyes,[1] J.S. Brooks,[1,2] † M.J.R. Hoch,[1]*

V. Srivastava,[3] R.D. James,[3] and C. Leighton[4]

[1]*National High Magnetic Field Laboratory, Florida State University, Tallahassee, Florida 32310, USA*

[2]*Department of Physics, Florida State University, Tallahassee, Florida 32310, USA*    (†*Deceased*)

[3]*Department of Aerospace Engineering and Mechanics, University of Minnesota, Minneapolis, Minnesota 55455, USA*

[4]*Department of Chemical Engineering and Materials Science, University of Minnesota, Minneapolis, Minnesota 55455, USA*



$Ni_{50-x}Co_xMn_{40}Sn_{10}$ shape memory alloys in the approximate range $5 \leq x \leq 10$ display desirable properties for applications as well as intriguing magnetism. These off-stoichiometric Heusler alloys undergo a martensitic phase transformation at a temperature $T_M$ of 300 - 400 K, from ferromagnetic (F) to non-ferromagnetic, with unusually low thermal hysteresis and a large change in magnetization. The low temperature magnetic structures in the martensitic phase of such alloys, which are distinctly inhomogeneous, are of great interest but are not well understood. Our present use of spin echo NMR, in the large hyperfine fields at $^{55}$Mn sites, provides compelling evidence that nanoscale magnetic phase separation into F and antiferromagnetic (AF) regions occurs below $T_M$ in alloys with $x$ in the range 0 to 7. At finite Co substitution the F regions are found to be of two distinct types, corresponding to high and low local concentrations of Co on Ni sites. Estimates of the size distributions of both the F and AF nanoregions have been made. At $x = 7$ the AF component is not long-range ordered, even below 4 K, and is quite different to the AF component found at $x = 0$; by $x = 14$ the F phase is completely dominant. Of particular interest, we find, for $x = 7$, that field cooling leads to dramatic changes in the AF regions. These findings provide insight into the origins of magnetic phase separation and superparamagnetism in these complex alloys, particularly their intrinsic exchange bias, which is of considerable current interest.


* Corresponding author: hoch@magnet.fsu.edu




# I. INTRODUCTION

Magnetic shape memory alloys of the type Ni-Mn-$X$ (where $X$ = Sn, In, Ga, *etc.*) have been the subject of intense recent investigation due to their technological importance and their rich variety of structural, mechanical, magnetic, and thermodynamic properties. Often these alloys undergo a martensitic phase transformation at or above room temperature, with great potential for applications in the areas of sensors, actuators, magnetic refrigeration, *etc.*[1-9] Much of the recent research on these materials has focused on off-stoichiometric versions of the full Heusler alloys, *i.e.* compositions of the form $Ni_{50}Mn_{25+y}X_{25-y}$, in which competing ferromagnetic (F) and antiferromagnetic (AF) interactions between Mn atoms located on inequivalent sites give rise to tunable magnetic phase competition. Substantial interest has arisen due to the fact that this phase competition appears to result in unanticipated magnetic behavior, including superparamagnetism (SP) in bulk samples, and intrinsic exchange bias (EB) effects.[10-12]

Important recent developments involve quaternary systems such as Ni-Co-Mn-$X$,[13-18] especially around the composition $Ni_{50-x}Co_xMn_{40}Sn_{10}$.[14-18] The rapid growth of interest in these Co-substituted alloys, and in particular those with $5 \leq x \leq 8$, is due in large part to the low thermal hysteresis at the martensitic transformation, which occurs at temperatures $T_M \approx 340$ K for $x = 6$. The low hysteresis is often valued in applications, as it minimizes the energy loss per cycle.[16] A marked change in the magnetic properties accompanies the martensitic transformation in these alloys, the transition from F order above $T_M$ to a state without long-range F order below $T_M$ being accompanied by a particularly large change in magnetization (approaching 1000 emu/cm$^3$,[16]). Notably, the development of these systems was guided to a significant extent by theoretical work predicting that thermal hysteresis at the transition is minimized in the presence of an invariant lattice plane at the martensite-austenite interfaces that occur in the transition region. This is explained by considering the ordered eigenvalues $\lambda_1$, $\lambda_2$, $\lambda_3$ of the structural transformation matrix $\mathbf{U}$, in particular requiring that to minimize hysteresis effects the determinant det $\mathbf{U} = \lambda_1\lambda_2\lambda_3 = 1$ with $\lambda_2 = 1$.[19, 20] This low hysteresis condition is found to hold to a good approximation in $Ni_{50-x}Co_xMn_{40}Sn_{10}$ for a relatively small range of $x$ values.[16-18] These alloys are therefore quite unique, with a relatively high Curie



temperature (up to 460 K), $T_M$ above ambient, and a martensitic transformation from a high magnetization F to a low magnetization non-F phase on cooling.

A magnetic phase diagram has recently been proposed by some of us for $Ni_{50-x}Co_xMn_{40}Sn_{10}$, with three regions, designated I ( $0 \leq x \leq 4.5$ ), II ( $4.5 < x \leq 11$ ) and III ( $x > 11$ ).[18] In this phase diagram, similar to an earlier one,[21] the valence electron per atom ratio, $e/a$, decreases with $x$ and there is an accompanying decrease in $T_M$, from 420 K at $x = 0$ to around 300 K at $x = 10$,[18] rapidly dropping to 0 K for $x \geq 11$. In terms of magnetism, in Region I the austenite phase is paramagnetic and the martensite phase exhibits no-long range F but shows distinct SP beneath a blocking temperature of ~100 K. In Region II ferromagnetism first emerges in the austenite (at $T_C$'s of 425-450 K), as the sample is cooled, followed by a martensitic transition that induces a transformation to a much lower magnetization non-F state. As in Region I, SP then emerges but now at lower temperatures ~50 K. The blocking temperature, $T_B$, thus exhibits a marked and largely unexplained change near $x = 4.5$, *i.e.* at the transition between Regions I and II. Finally, in Region III, the F austenite phase persists from $T_C$ ( $\leq 450$ K) to the lowest temperatures, with no martensitic transformation. Notably, the EB blocking temperature, $T_{EB}$, below which intrinsic EB effects are detected, is distinct from $T_B$ in Region I but coincident with $T_B$ in Region II. The range $5 \leq x \leq 8$, in Region II, has been suggested as optimal for magnetic shape memory and other applications.[16,17]

Recent work has shown that the unexpected SP and EB effects are linked to the formation of a magnetically inhomogeneous low temperature state in these alloys, as seen in small-angle neutron scattering (SANS) experiments on polycrystalline $Ni_{50-x}Co_xMn_{40}Sn_{10}$ with $x = 6$ and 8, made in the temperature range $30 - 500$ K. The results provide strong evidence of nanoscale magnetic phase separation in this material, with a liquid-like spatial distribution of SP clusters forming in a non-F matrix.[17] Although AF order and paramagnetism in the matrix could not be distinguished, a number of indirect observations pointed to some form of AF order.[17,18] Magnetization measurements, made as a function of temperature, provide additional evidence for clusters, and in the case of $x = 6$ give $T_B \approx 60$ K.[17] The SANS results, together with magnetization findings, provide information on the F cluster diameters,



estimated to be in the range 2 – 6 nm, with mean center-to-center separation 12 nm. However, the detailed nature of the F clusters, and their size distribution, remain unclear and, in particular, the characteristics of the paramagnetic, or AF, matrix regions have not been determined. Many open questions concerning the magneto-electronic phase separation in the martensitic phase of the $Ni_{50-x}Co_xMn_{40}Sn_{10}$ alloys therefore remain to be addressed.

Our recent work, using $^{55}Mn$ NMR on $Ni_{50}Mn_{40}Sn_{10}$ (*i.e.*, $x = 0$), sheds further light on these issues by investigating the SP at this composition, and, critically, by elucidating the nature of the non-F matrix phase.[22] The latter was shown to indeed be AF as opposed to simply paramagnetic, but with short- rather than long-range magnetic order. In essence both F and AF clusters occur, with both a distribution in sizes, and distinct temperature dependences. In the present work we again apply low temperature $^{55}Mn$ NMR, this time studying polycrystalline $Ni_{50-x}Co_xMn_{40}Sn_{10}$ across all three interesting magnetic regions, specifically at $x = 7$ (in the martensitic phase) and $x = 14$ (in the austenite phase), for comparison with our previous findings for $x = 0$. The $^{55}Mn$ hyperfine-field-NMR results provide detailed information on the ground state magnetic properties as a function of $x$, and on the evolution of the inhomogeneous nanoscale magnetic structures with temperature. At $x = 7$ two distinct F regions and a dominant AF region are detected at low temperatures, with the F regions having very different characters (identified with high and low local concentrations of Co), and the AF regions being highly thermally unstable. This is quite different to $x = 0$, reflecting the increase in stability of the F phase with Co substitution, confirmed by our measurements at $x = 14$. Information obtained on the size distribution of the nanoscale F and AF clusters, together with their proximity and temperature dependent dynamics, helps to elucidate the origins of the macroscopic magnetic properties represented in the phase diagram of these alloys. In particular, radical changes in AF regions are evidenced with field cooling at $x = 7$, directly relevant to the intrinsic EB phenomena that have attracted such attention in these and related systems.



## II. EXPERIMENTAL RESULTS

Polycrystalline samples of $Ni_{50-x}Co_xMn_{40}Sn_{10}$ with selected $x$ (0, 7 and 14) were prepared by arc melting high purity metal starting materials under argon followed by annealing at 900 °C for 24 hours before quenching in water.[16-18] The ingots were then powdered and annealed to provide samples for NMR (with suitable RF penetration), as described previously.[22] Characterization was carried out using energy dispersive spectroscopy, differential scanning calorimetry, X-ray diffraction, and magnetometry. Experiments on the $x = 0$ sample are described in our prior work.[22] At $x = 7$ the transition temperatures are $T_M = 370$ K and $T_C = 430$ K, with low thermal hysteresis at the martensitic transition, $\Delta T_M \sim 6$ K. The $x = 14$ sample has $T_C = 460$ K and does not undergo a martensitic transformation.

The $^{55}Mn$ (gyromagnetic ratio $\gamma/2\pi = 10.5$ MHz/T) NMR measurements presented here were made mainly in zero applied magnetic field but also in fields, $\mu_0 H$, up to 8 T using a spin echo spectrometer with automated frequency sweep capability over the range $f = 150 - 450$ MHz. The polycrystalline metallic grains were embedded in paraffin wax to minimize skin depth effects. Zero-field (ZF) spectra were observed in the hyperfine fields experienced by $^{55}Mn$ nuclei, which ranged from ~20 to 43 T, and were recorded as a function of temperature ($T$) by increasing the frequency in 1 MHz steps over the required range. As described in Ref. 22, the magnetic system NMR signal enhancement factor, $\eta$, which is given by $\eta = B_{hf}/B_A$, with $B_{hf}$ the hyperfine field and $B_A$ the anisotropy field, can effectively distinguish F and AF regions since $\eta$ is, typically, significantly larger in F regions than in AF regions. As discussed in our prior work,[22] further confirmation of the F / AF assignments is provided by measurements of spectral frequency shifts in low (~ 1T) applied magnetic fields; lines due to the F component downshifts with $H$, in contrast to the behavior of spectral lines from AF regions.

Low temperature (1.6 K), zero-field-cooled (ZFC) NMR spectra, recorded at optimal RF power levels in each case, and obtained by integrating the spin echo signals at stepped frequencies, are shown in Fig. 1 for $x = 0$, 7 and 14. The $x = 0$ spectrum has previously been discussed,[22] and is shown for comparison



with the $x = 7$ and $x = 14$ spectra. On this plot the open circle (blue) data are associated with low $\eta$ ($\sim 10$) spectral components from $^{55}$Mn sites in regions of the sample that have AF character, and are labeled "AF". In contrast, the open circle (red) data are due to F regions, with $\eta \sim 225$, and are labeled "F". Comparison of the spectra shows that substitution of Co on a fraction of the Ni sites leads to major changes in the spectral features. In particular, the $x = 7$ and $x = 14$ spectra have F components at $f > 350$ MHz, designated Region H in the figure (high frequency), that are not present for $x = 0$. In addition, for $x = 7$ the spectral features below 350 MHz, designated Region L in the figure (low frequency), involve broad AF and F components, with the AF of dominant importance only at the lowest temperatures ($T < 5$ K, see below). Note that for $x = 7$ the amplitude of the large AF component has been reduced by a factor 10. A comparison of the spectral areas at this $x$ value shows that at 1.6 K the AF constitutes $\sim 90$ % of the magnetic material in this sample as discussed in greater detail in Section IIIA. For $x = 14$ on the other hand, no AF signals are observed over the entire frequency range but there are distinct F features in Region H and a broad weak response in region L. These changes in the spectra induced by the substitution of Co on some of the Ni sites are discussed in full in Section III, but we note immediately that they clearly reflect radical changes in the magnetic properties.

As the temperature is raised the forms of the F spectral features for $x = 7$ and $x = 14$ do not change significantly, but the amplitudes steadily decrease and the 1.6 K spectral peaks in Region H gradually shift to lower frequencies. Fig. 2 illustrates the latter point by showing the $T$ dependence of the frequency ratio $f / f_0$ for the principal F spectral component for $x = 7$, taking $f_0 = 435$ MHz at 1.6 K. Over the range 1.6 to 150 K $f$ decreases from 435 MHz to 405 MHz. Qualitatively similar shifts are found over this $T$ range for the F component at $f_0 \sim 400$ MHz (see Fig. 1). Notably, the frequency ratio $vs$. $T$ plot in Fig. 2 shows a marked decrease in $f / f_0$ as $T$ increases from 1.6 to 5 K (inset), followed by a more gradual linear decrease from 5 to 70 K. For $T > 70$ K the ratio $f / f_0$ decreases rapidly again, and the NMR signal amplitude diminishes to zero. The approximately linear $T$-dependence (solid black fit line) suggests a mechanism involving dynamics of the F clusters, as discussed below, and established in our prior work.[22]



In order to also analyze the evolution of the spectral weight with temperature, the peaks for the $x = 7$ and $x = 14$ samples were fit with multiple Gaussian curves (not shown). For $x = 7$, the behavior of the Curie-law-corrected area of the F components in Region H is plotted (in arbitrary units) as a function of $T$ in Fig. 3. This spectral area is proportional to the number of $^{55}$Mn nuclei which experience hyperfine fields that are effectively *static* on the NMR timescale (~10 $\mu$s) and which can therefore contribute to the observed spin echo signals.[22] The temperature scaling compensates for the Curie law decrease in the nuclear magnetization. For SP clusters of volume $V$ the NMR blocking temperature, $T_B^{NMR}(V)$, is the temperature above which NMR signals from these clusters can no longer be detected.[22] If there is a distribution of cluster sizes it follows that there is a corresponding distribution of $T_B^{NMR}$ values.

It can be seen from Fig. 3 that the $T$-scaled integrated area (in arbitrary units) of the F component at $x = 7$ plateaus at low $T$, but exhibits a gradual fall-off with increasing $T$, up to 300 K. The F spectral area *vs.* $T$ plot in Fig. 3 is fit (solid red line) using the SP cluster model we previously established,[22] with a distribution of F cluster sizes (inset), as described further in Section III. In marked contrast, the dashed line in Fig. 3 shows the rapid decrease in the $T$-corrected areas of the AF component in Region L (again in arbitrary units to facilitate a comparison with the behavior of the relatively small F component). The AF shows no plateau down to 1.6 K, very different to the $x = 0$ case in Ref. 22. The rapid decrease in amplitude of the Region L broad AF component with increasing $T$, until it is no longer observable at $T >$ 10 K, evidences a marked decrease in thermal stability of the AF regions in comparison to the AF at $x =$ 0. It should be noted, however, that at $x = 7$ the broad low frequency F component (in Region L) is still observable, albeit with reduced amplitude, above 100 K. At low $T$ ($< 4$ K), and in low $H$, this F component is masked by the dominant AF, and emerges in the spectra only above 10 K.

Fig. 4 highlights a different aspect of the data. Specifically, Fig. 4(a) shows the $T$ dependence of the $x = 7$ spin-lattice relaxation time $T_1$ for the principal high frequency F component, while Fig. 4(c) gives the behavior of $T_1$ for the AF component in Region L. Measurements were made in zero field (ZF) and in 1 T.



Figures 4(b) and 4(d) then show the corresponding Korringa plots of $1/T_1T$ $vs.$ $T$ for the F (Region H) and AF (Region L) components, respectively. These dynamic data are discussed in full in Section III.

Finally, the transverse or spin-spin relaxation time $T_2$, obtained from stretched exponential fits to the spin echo decay curves at $x = 7$, is found to be weakly $T$ dependent, and, for the F components in Region H, decreases from 110 $\mu$s to <10 $\mu$s as $T$ increases from 1.6 K to 120 K. Fig. 3 shows that the *spectral area* decreases steadily over this same temperature range, as the regions in which $T_2 < 10$ $\mu$s no longer contribute significantly to spin echoes. $T_1$ in the F regions is roughly an order of magnitude longer than the corresponding $T_2$.

### III. DISCUSSION

The principal findings of the present work may be summarized as follows. Substitution of Co for Ni in $Ni_{50-x}Co_xMn_{40}Sn_{10}$ leads to dramatic changes in the low temperature $^{55}$Mn ZF spectra for $x = 7$ and $x = 14$, in comparison with the $x = 0$ spectrum, as can be seen in Fig. 1. Among other differences, new F components appear above 350 MHz. This major change in the form of the spectra with $x$ suggests that Co substitution for Ni produces significant changes in the local electronic structure, generating enhanced hyperfine fields at $^{55}$Mn sites in the vicinity of the Co ions. In addition, at $x = 7$ the broad F and AF spectral components in Region L, centered at about 275 MHz and 230 MHz, respectively, occur at frequencies similar to those of the spectral features at $x = 0$. However, the narrow F and AF components seen near 300 and 250 MHz for $x = 0$,[22] are *not* seen at $x = 7$. In our prior work on $x = 0$ [22] these features were interpreted as originating from the largest F and AF clusters in the sample. The $x = 7$ F and AF spectral lines in Region L are attributed to Mn in regions that have comparatively low Co concentrations and correspondingly little change in the $^{55}$Mn hyperfine field compared to $x = 0$. Below, we provide further discussion of specific aspects of the data.



## A. NMR spectral amplitude changes with temperature

The Curie law (*T*-scaled) corrected spectral areas (proportional to the number of $^{55}$Mn nuclei in regions that contribute to the corresponding NMR signals) steadily decrease with *T*, consistent with a distribution of cluster sizes, and a corresponding distribution of $T_B^{NMR}(V)$ values.[22] Most notably, the area of the 230 MHz AF component decreases dramatically in the range 1.6 – 10 K, as shown in Fig. 3 for *x* = 7, evidencing thermal instability of these regions with low $T_B^{NMR}$ characteristics. In contrast, the *T*-scaled area of the broad low amplitude F spectral component in Region L decreases gradually with rising *T* from 1.6 – 150 K, similar to that of the 435 MHz component shown in Fig. 3. This behavior points to broad distributions of the F cluster sizes and associated NMR blocking temperatures in both Regions L and H.

The spectral area results, considered together, suggest that the largest F and AF ordered regions, responsible for the narrow spectral features found below 350 MHz for *x* = 0,[22] have their long range correlated spin structures distinctly altered by the incorporation of Co on Ni sites in the *x* = 7 and *x* = 14 cases. This important finding indicates that the abrupt decrease in the DC magnetization SP blocking temperature, $T_B$, from ∼ 120 K to 70 K, at around *x* = 4.5 in the phase diagram,[18] *i.e.* the Region I/Region II boundary, which has proven difficult to understand, is a result of a reduction in the size of SP clusters produced by the substitution of Co on a "critical" fraction of Ni sites. Furthermore, it is possible that local austenite-like regions are produced at sufficiently large Co concentrations. While these arguments in general agree with prior statements on decreasing thermal stability of the F clusters in Region II of the phase diagram compared to Region I (consistent with the reduced $T_B$),[18] this new understanding highlights changes in the cluster size *distribution* (*i.e.* elimination of the largest clusters), which was not explicitly considered earlier.[18] As mentioned above, the substitution of Co for Ni results in a large change in $B_{hf}$ at $^{55}$Mn sites, indicating that the local electronic structure is considerably modified by the Co substitution. While it is possible that there is a transfer of spin density to Mn produced by the replacement of Ni by Co this is likely to be a small effect as the bulk of the spin density is already concentrated on the



Mn in the $Ni_2MnSn$ system.[23] The local crystal structure in a particular region is likely to depend on the Co concentration and the geometrical configuration of Co on neighboring Ni sites. Variations in this local structure could produce the multiple components seen in Fig. 1 for $x = 7$ and 14, together with the observed increase in linewidth as $f$ decreases. It is also interesting to note the similarity in the high frequency, Region H, spectral lineshapes in the $x = 7$ and $x = 14$ alloys. This similarity supports the suggestion that for $x = 7$ the austenite phase is stabilized at low temperatures in regions with high Co concentrations.

In order to quantitatively compare the relative concentrations of Mn in the various magnetic regions of the present samples it is necessary to take into account variations in the RF enhancement factor, $\eta$, which was used in Section II to distinguish the AF and F components in the $x = 7$ $^{55}$Mn spectra. Using the values for $\eta$ given in Section II, together with the relationship $\eta = B_{hf}/B_A$ yields $B_A = 0.19$ T for the anisotropy field of the F component in Region H for $x = 7$ with a similar value for $x = 14$. This value for the anisotropy field is similar to that found for the F components in $Ni_{50}Mn_{40}Sn_{10}$.[22] From the relative areas of the $^{55}$Mn spectra, which are shown with proper allowance for the difference in $\eta$ values, it is then possible to obtain the ratio of the numbers of F and AF Mn atoms, denoted $N_F$ and $N_{AF}$ respectively. For $x = 7$ at 1.6 K we obtain $N_F/N_{AF} = 0.1$, meaning that the AF component is of dominant importance, producing ~ 90 % of the combined F plus AF NMR signal. This is in contrast to the $x = 0$ case, in which $N_F/N_{AF} \approx 0.8$ $i.e.$ close to unity.[22] The major part of the F constituent (~75 %) for $x = 7$ is in Region H.

The substitution of a fraction of the Ni by Co leads to major changes in the nanoscale magnetic structure of these systems. In a comparison of the NMR results obtained for $x = 0$ and $x = 7$ it is the disappearance of *spatially extended* AF regions for $x = 7$, as revealed by their low $T_B^{NMR}$ values (< 10 K), as inferred from Fig. 3, which is particularly striking. In addition, the NMR-detected F regions at $x = 7$ are separated into the low and high frequency components, in Regions L and H respectively, and, as discussed above, are, at 1.6 K, greatly diminished in relative intensity compared to AF. These findings suggest that in



regions of these alloys the exchange interactions between Mn atoms are significantly altered through the replacement of some Ni by Co. Note as an aside that the NMR results suggest that a fraction of the smallest F and AF clusters are not detected, even at temperatures as low as 1.6 K, due to cluster dynamics.

### B. Cluster size estimates

In order to quantify the cluster sizes, especially for the F regions generating the SP behavior, analysis of the data shown in Fig. 2 is useful. In particular, The SP globular cluster model described in Ref. 22 predicts that for a F cluster of volume $V$ at temperatures below $T_B^{NMR}(V)$ the slope of an $f/f_0$ vs. $T$ plot is given by $k_B/(2K_AV)$ where $K_A$ is the anisotropy energy density. For a system in which there is a distribution of cluster sizes the analysis is less straightforward. In order to obtain a rough estimate of the average cluster size, however, we assume that the cluster model can be applied to the $f_0 = 435$ MHz component frequency dependence on $T$ (Fig. 2). We take $K_A = \frac{1}{2}n\mu B_A$ where $\mu = 4\mu_B$, $n$ is the spin density and $B_A = 0.19$ T, as given above. (We assume that $K_A$ is the same for all the F clusters but we cannot exclude the possibility of some distribution of $K_A$ values linked to cluster sizes and shapes.) From the slope of the straight line fit through the points in the range $3 - 70$ K in Fig. 2 we obtain an average cluster diameter $D \sim 9$ nm. It should be noted here that because of the distribution of cluster sizes, with a corresponding distribution of $T_B^{NMR}(V)$ values, there is a gradual decrease in signal amplitude with $T$ (see Fig. 3) as the smaller clusters pass through their respective blocking temperatures. It therefore follows that this size estimate (9 nm) is somewhat skewed towards the larger clusters. The inset to Fig. 2 shows that at the very lowest $T$ (1.6 – 3 K) the slope of the $f$-$T$ curve increases, consistent with a growth in proportion of small volume clusters, accompanied by an upward frequency shift. The size estimate of 9 nm from the slope in Fig. 2 is somewhat larger than the previously reported value of 2 - 6 nm for $x = 6$,[17] a point that will be returned to below. As a final comment on the analysis of Fig. 2 we note that above 70



K the frequency variation is no longer linear in $T$ and the simple form given by the cluster model no longer applies.

It is possible to obtain further information on the size distribution of the F clusters in Region H which, based on the 1.6 K spectral areas, make up ~ 7 % of the total NMR detected magnetic material (with AF ~90% and total F ~10%) in the $x = 7$ sample at low $T$ by using our SP cluster model to fit the Curie-law-corrected area plot in Fig. 3. The procedure is described in Ref. 22 and involves the use of the following expression for the nuclear spin-spin relaxation rate for a globular cluster of volume $V$

$$\frac{1}{T_2} = \tfrac{1}{12} S(S+1)\left(\frac{\omega_I^2}{\omega_S^2}\right)\left(\frac{T}{T^*}\right)^2\left(\frac{1}{\tau_c}\right),$$  (1)

where $\omega_I = 2\pi f_{hf}$, $\omega_S = 2\pi/\tau_0$ with $\tau_0$ the pre-exponential factor in the Néel–Arrhenius (N-A) expression, and $T^* = K_A V/k_B$. The correlation time, $\tau_c$, is determined by the coupling of a cluster spin to the lattice and can, in principle, be modeled using a modified form of the N-A relation with the anisotropy energy reduced by a significant factor. For clusters of volume $V$, with the minimum observable $T_2 \sim 10$ $\mu$s, we find that $\tau_c \to \tau_0$ as $T \to T_B^{NMR}(V)$. For a system in which there is a distribution of SP cluster volumes, described by a probability distribution $P(V)$, there is a corresponding distribution of $T_2$.[22] The amplitude of the NMR signal will then decrease as $T$ is raised and fewer clusters contribute to the signal as their $T_2$ values fall below the NMR measurement time $\tau_m = 10\ \mu$s. Making use of the Poisson distribution, the model predicts that the NMR spin echo signals from clusters of a particular size should decay as $S(T) = S_0 \exp\left(-\tau_m/T_2\right)$, where $S(T)$ is the signal amplitude at temperature $T$ and $S_0$ is the signal amplitude from these clusters at the lowest $T$. Using Eq. (1) and integrating numerically over a log-normal volume distribution, provides the reasonable fit to the F spectral area plot shown in Fig. 3. The size distribution of the F clusters obtained from the fit is shown in the inset to Fig. 3, the most probable diameter being $D \sim 4.4$ nm. We emphasize that while the fitting process does involve a number of



adjustable parameters, the size distribution is found to be fairly robust to changes in these parameters. However, the integration over cluster volumes using a broad size distribution may be an oversimplification of the actual situation. Comparison of the sizes obtained from the two approaches we have used, based on Figs. 2 and 3, respectively, which differ by a factor 2, suggests that the average cluster diameter lies in the range 3 - 10 nm with the most probable diameter ~ 6 nm. Notably, this range overlaps the estimate of 2 - 6 nm made previously for $x = 6$.[17] Note that the line for the AF spectral area curve in Fig. 3, over the limited range 1.6 – 10 K, is simply a guide to the eye, rather than a fit. The nature and size of the AF regions is unclear but the NMR evidence suggests numerous small regions with $D \sim 1$ nm but which are probably not globular in shape.

While the spheroidal cluster model approach used in the analysis of the NMR results is certainly an idealization of the magnetic substructures in $Ni_{50-x}Co_xMn_{40}Sn_{10}$, the cluster size estimates thus obtained are consistent with the SANS and magnetometry estimates.[18] The microscopic picture of the $x = 7$ alloy that is provided by the $^{55}Mn$ NMR results is summarized as follows: There are two different types of F regions, corresponding to Co-rich and Co-poor clusters, respectively. These F regions have a distribution of nanoscale sizes, and, in addition, there is an AF component which is of dominant importance in the system at temperatures below 10 K. Above 10 K, the AF region cannot be detected by spin echo NMR due to collective spin dynamic effects which shorten the transverse (spin-spin) nuclear relaxation times. The F components associated with Co rich and Co poor regions are detected up to 150 K, but both signals show a marked decrease in amplitude with rising temperature as a result of a cluster size distribution in the range 3 – 10 nm. At $x = 14$ no AF is detected while two F components are found with the Region H response from Co-rich regions being of dominant importance.

### C. Connections to exchange bias effects

Having elucidated the $x$ dependence of the low $T$ spectra shown in Fig. 1, and having quantitatively analyzed the behavior shown in Figs. 2 and 3 to extract volume fractions of the coexisting AF and F



phases, relative thermal stabilities, and cluster sizes, we now turn specifically to the EB effects that have gathered such attention in these and related materials. As mentioned in Section I, EB effects are detected in DC magnetization measurements on $x = 7$ samples, for instance, at $T < T_B \sim 50$ K. For EB to occur , in a simple picture, it is necessary that interactions between spins at F-AF interfaces provide a pinning mechanism for the F spins.[24-27] In experiments involving layered magnetic films it has been shown (e.g. Refs. 27, 28) that uncompensated spins in the AF play a crucial role in establishing EB effects.

In order to investigate the EB in detail we carried out field cooling (FC) experiments on the $x = 7$ alloy. Figure 5(a) shows the Region L AF spectral component, centered at 240 MHz, in ZF following FC to 1.7 K with fields in the range $0 \leq \mu_0 H \leq 6$T. Surprisingly, striking changes in the amplitude of the AF component are produced by FC. The 255 MHz F component in Region L is not detectably changed by FC for $\mu_0 H \leq 0.15$ T, while at higher fields the behavior of this component is masked by the large growth in amplitude of the AF component. On the other hand the F components in Region H, which comprise $\sim 70$ % of the total F regions, are unchanged in FC-ZF-run experiments, as can be seen in Fig. 5(b). The FC response of the AF and F components at this $x$ value are thus starkly different. The increase in AF spectral amplitude is in part attributed to alignment of spins along $H$ (perpendicular to the RF field), which is the optimal orientation for NMR, but more significantly is due to an order of magnitude *increase* in $\eta$ as shown in Fig. 5(d). This figure panel shows $\eta$ as a function of cooling field for the AF/ Region L, F/ Region L, and F/ Region H components, emphasizing that only the AF is strongly influenced by field cooling. These $\eta$ values are derived from  Fig. 5(c), which shows 1.7 K plots of the ZF spin-echo amplitudes, at selected frequencies (240 and 275 MHz), following FC, recorded as a function of the amplitude of the RF field $H_1$ for a fixed RF pulse length. It is clear that following FC the optimal $H_1$ shifts to lower values, corresponding to an increase in $\eta$ . Note that for $f = 275$ MHz at low $\mu_0 H$ (< 0.1 T) there are two distinct maxima in the spin-echo response curves, corresponding to the high $H_1$ , low $\eta$ , AF components, and the low $H_1$, high $\eta$ , F components, respectively. As noted above, for $f = 240$ MHz the F



component is less well resolved, and the AF is of dominant importance. As $\mu_0 H$ is increased the difference in the $\eta$ values for the AF and F components decreases, as seen in Figs. 5(c) and 5(d). Interestingly, in FC experiments for $x = 0$ the observed changes in the signal amplitude and in $\eta$ are much smaller than in $x = 7$ where the AF and F cluster sizes are reduced in size. The enhanced detection of EB-linked effects by NMR for $x = 7$ compared to $x = 0$ is clearly facilitated by modifications in the nanoscale geometry of spin clusters induced by the insertion of Co for Ni.

Based on these observations it is possible to reach the following conclusions on EB in the $x = 7$ system. First, the dominant F components in Region H, which are associated with high Co concentrations, and modified local crystal structures, appear to play, at best, a minor role in the EB process, while the interactions between the F and AF components in Region L are of dominant importance in producing the effect. Second, exchange interactions at the AF-F interfaces in Region L produce an increase in the $\eta$ values for the AF of more than an order of magnitude. These $\eta$ values approach those of the F regions to within a factor ~3. This novel and interesting finding concerning the coupling of spins, and their linked dynamics at the F-AF interface in $x = 7$, complements previous EB experiments involving magnetization measurements on $Ni_{44}Co_6Mn_{40}Sn_{10}$.[17,18] In addition, the change in the nanoscale cluster properties of the AF component between $x = 0$ and $x = 7$ provides a basis for explaining the distinct change in $T_{EB}$ at $x \sim 4.5$ in the phase diagram given in Ref. 18.

As the temperature is raised towards 10 K the role of the AF component, as detected by NMR, diminishes in importance (Fig. 3). The DC magnetization measurements show that for $4.5 \leq x \leq 10$ EB effects are no longer observable above 50 K which is also the DC SP blocking temperature. The significant difference in $T_{EB}$ from NMR (~10 K) and DC measurements (50 K) reflects the different response times of these techniques to collective spin dynamics. Differences between NMR $T_B^{NMR}$ and DC $T_B$ values are, as with EB effects, attributed to the different dynamic processes detected by the two types of measurements, as discussed in Ref. 22. In sum, these low temperature experiments show that for $x = 7$ the



AF/Region L spin clusters or layers, seen below 10 K in Fig. 3, are of dominant importance in providing the EB interfacial pinning mechanism for the F/Region L spins. NMR thus provides compelling evidence that the shift in the DC $M - \mu_0 H$ loop due to EB is primarily due to interactions at the AF-F interfaces in Region L associated with the low Co martensitic components. In addition, interactions between F spins in Regions L and H are likely to play some role in the EB process.

### D. Spin-lattice relaxation and cluster dynamics

Turning now to further analysis of dynamic data, we note that the $T_1$ vs. $T$ plots for the high frequency F component in ZF and in 1 T, as shown in Fig. 4 (a) for $x = 7$, reveal an initial marked decrease below 20 K, followed by a more gradual decrease at higher $T$. For the AF component, $T_1$ decreases rapidly with increasing $T$, over the range 1.6 – 10 K, consistent with thermally induced small cluster spin flips which give rise to fluctuating local fields responsible for spin-lattice relaxation. The most dynamic small clusters do not contribute to the spin echo signals, since $T_B^{NMR} < 1.6$ K for these clusters. They can, of course, play a role in inducing relaxation in their larger neighbors which are observed by NMR.

We have not attempted a quantitative analysis of the AF relaxation behavior shown in Figs. 4(c) and (d) and concentrate on the F component $T_1$ in Region H over the range 1.6 – 150 K, as presented in Figs. 4(a) and (b). In order to explain the behavior of $T_1$ with $T$ in Fig. 4(a) two competing relaxation mechanisms need to be considered: Firstly, inter-cluster dynamic dipolar interactions, involving the small cluster spin flips, and secondly, the intra-cluster Korringa mechanism involving conduction electron scattering. The relative importance of these mechanisms is dependent on the cluster size and it is instructive to consider the Korringa-style plot of $1/T_1 T$ *vs.* $T$ in Fig. 4(b). The large decrease in $1/T_1 T$ produced by a field of 1 T points to field-suppression of the fluctuating inter-cluster dipolar mechanism through alignment of cluster moments parallel to $H$ and a reduction in the transverse fluctuating field because of partial cancellation of fields from various neighbor clusters. The Korringa plateau value for $T > 20$ K in Fig. 4(b) then allows us to estimate the density of states at the Fermi level in the larger F clusters using the Moriya expression



$$\frac{1}{T_1^K} = C \left\langle \frac{1}{r_A^3} \right\rangle_F \left( \rho_\uparrow^2 + \rho_\downarrow^2 \right) F\left( \Gamma \right)$$, where $C = \left( 16\pi/5 \right) \mu_0^2 \gamma_I^2 \mu_B^2 \hbar k_B T$ with $\mu_B$ the Bohr magneton, $\gamma_I$ the

nuclear gyromagnetic ratio and $\rho_\uparrow$ and $\rho_\downarrow$ the densities of states (DOS) at the Fermi energy, $E_F$, for spin

up and spin down electrons respectively.[29] The inverse radius cubed, averaged over the Fermi surface, is

$\left\langle 1/r_A^3 \right\rangle_F = 0.32$ m$^{-3}$.[22] The function $F\left( \Gamma \right)$ gives the $t_{2g}$ orbital admixture in the wave function at $E_F$. If we

take $F\left( \Gamma \right) = 1$ as an approximation in estimating the DOS at the Fermi energy we obtain

$\left( \rho_\uparrow^2 + \rho_\downarrow^2 \right) = 0.8$ eV/ f.u. which is roughly half the value obtained for the DOS in F regions of

Ni$_{50}$Mn$_{40}$Sn$_{10}$.[22] Substitution of Ni by Co leads to this large decrease in the Fermi level DOS for the 435

MHz F component (Region H) in the $x = 7$ alloy.

## IV. CONCLUSIONS

Our hyperfine-field-mediated NMR measurements provide considerable nanoscale information on the

magnetic nature of the shape memory alloys Ni$_{50-x}$Co$_x$Mn$_{40}$Sn$_{10}$ with $x = 0$, 7, and 14. In both finite $x$

alloys the incorporation of Co on a fraction of the Ni sites leads to the formation of new types of F

nanoregions, or clusters, with distinct NMR features when compared to $x = 0$. In the case of $x = 14$ at low

temperatures the spectral signatures of two F regions are identified (although one is dominant), and it is

suggested that these regions correspond to high and low local Co concentrations. For $x = 7$, at

temperatures well below the martensitic transition, there are also two F regions, again linked to high and

low Co, this time coexisting with an AF region. Both the F regions have broad size distributions, as

revealed by spectral area versus temperature plots over the range 1.6 – 200 K. The results have been

interpreted in terms of a distribution of NMR blocking temperatures in each case. The similar mixed

magnetic phase behavior found for both $x = 7$ and $x = 14$ suggests that local structural variations persist to

low temperatures in these systems, possibly linked to the coexistence of small austenite and martensite



regions. Signals from the AF regions, which play the key role in establishing intrinsic exchange bias effects in the $x = 7$ system, are observed in NMR spin echo experiments only at temperatures below 10 K. This thermally unstable AF component, which may be viewed as a matrix in which the F regions are embedded, is of dominant importance in terms of volume fraction at temperature below 5 K. These regions provide a dynamically correlated environment for the F clusters as the temperature is raised. Field-cooled experiments show that it is the low Co concentration F regions that couple to the AF component to establish EB. Direct evidence for this interaction is obtained from changes in NMR signal enhancement.

**Acknowledgments**

The work at the National High Magnetic Field Laboratory was supported by NSF DMR-1157490 and by the State of Florida. JSB received funding from NSF-DMR 1309146. Work at UMN in CL's group supported by DOE under award DE-FG02-06ER46275. Work at UMN in RDJ's group supported by AFOSR-MURI (FA9550-12-1-0458), NSF-PIRE (OISE-0967140) and ONR (N00014-14-1-0714). The assistance of Daniel Phelan with sample fabrication is gratefully acknowledged.

**Figure captions**

**FIG. 1 (color online)**. Frequency ($f$) scanned $^{55}$Mn zero applied field NMR spectra in polycrystalline Ni$_{50-x}$Co$_x$Mn$_{40}$Sn$_{10}$ at 1.6 K for $x = 0$, 7 and 14. Spectra associated with ferromagnetic regions (F, red) and antiferromagnetic regions (AF, blue) are indicated. The amplitude of the dominant AF component has been reduced by a factor 10 in the $x = 7$ spectrum. Comparison of the spectra reveals that substitution of Co on Ni sites at $x = 7$ and 14 produces ferromagnetic regions or clusters with enhanced hyperfine fields, and spectral features above 350 MHz, compared to $x = 0$. The spectra can be well fit using multiple Gaussian curves. The spectral region below 350 MHz is designated "L" (low frequency) and the region above 350 MHz is designated "H" (high frequency), as discussed in the text.

**FIG. 2 (color online)**. The ratio, $f/f_0$, of the $^{55}$Mn NMR resonance peak center frequency ($f$) to the resonance frequency ($f_0$) at 1.6 K, as a function of $T$, for the principal F component in Region H (*i.e.* the highest frequency in Fig.1) for Ni$_{43}$Co$_7$Mn$_{40}$Sn$_{10}$. The frequency decreases approximately linearly at temperatures in the range 3 – 70 K and then more rapidly at higher temperatures. The decrease in frequency is produced by averaging of the hyperfine field by collective cluster spin fluctuations in their anisotropy fields. The slope of the straight line fit through the points provides an estimate of the average cluster size, as discussed in the text. The small reproducible increase in $f/f_0$ at the lowest temperatures (< 3K), as shown in the inset, is attributed to freezing of very small clusters and a possible change in the spectral lineshape.

**FIG. 3 (color online)**. Curie-law-corrected $^{55}$Mn NMR peak areas for the high frequency (region H in Fig. 1) ferromagnetic components (black circles) in Ni$_{43}$Co$_7$Mn$_{40}$Sn$_{10}$ as a function of temperature. The fitted curve is based on a cluster dynamics model, with a distribution of cluster sizes, as described in the text. The inset shows the log-normal cluster radius distribution used in the fit; assuming spherical clusters



the most probable diameter is 4.4 nm. For comparison the Curie-law-corrected areas for the AF

component (region L in Fig. 1) are also shown (blue triangles). This AF component cannot be detected at

$T > 10$ K.

**FIG. 4 (color online).** [55]Mn spin-lattice relaxation times $T_1$ for (a) the principal F spectral component in

Region H, and (c) the AF components in Region L (see Fig. 1) as a function of temperature in

$Ni_{43}Co_7Mn_{40}Sn_{10}$. Measurements were made in both zero field and in an applied field of 1 T. Comparison

of the F and AF $T_1$ values shows that the AF relaxation times are substantially shorter than those of the F

component, and that they decrease more rapidly as $T$ increases. The field dependence seen in Fig. 4(a) is

discussed in the text. Figures (b) and (d) are the Korringa plots of $1/T_1 T$ $vs.$ $T$ for the F and AF

components, respectively. An estimate of the density of states at the Fermi level for the F components is

given in the text.

**FIG. 5 (color online)**. (a) [55]Mn zero field spectra in the range $150 - 375$ MHz, obtained in various RF

fields, $H_1$, optimized for the 240 MHz AF component, following field cooling to 1.7 K in varied applied

fields up to 6 T. (b) [55]Mn zero field spectra, obtained in a fixed low RF field optimized for the 435 MHz

component, over the range $150 - 500$ MHz, following field cooling to 1.7 K in fields up to 6 T. (c) Spin

echo amplitude response curves at 275 MHz and 1.7 K obtained as a function of $H_1$ following zero field

cooling, in fields up to 6 T. The marked *decrease* in the optimal $H_1$ with increase in the $\mu_0 H$ used during

field cooling, is due to an unexpected large increase in the enhancement factor $\eta$ as discussed in the text.

(d) Plot of $\eta$ vs. $\mu_0 H$ following field cooling to 1.7 K, as obtained from Fig. 5 (c). The dramatic increase

in $\eta$ is attributed to exchange bias linked interactions between the F and AF regions across their

interfaces.



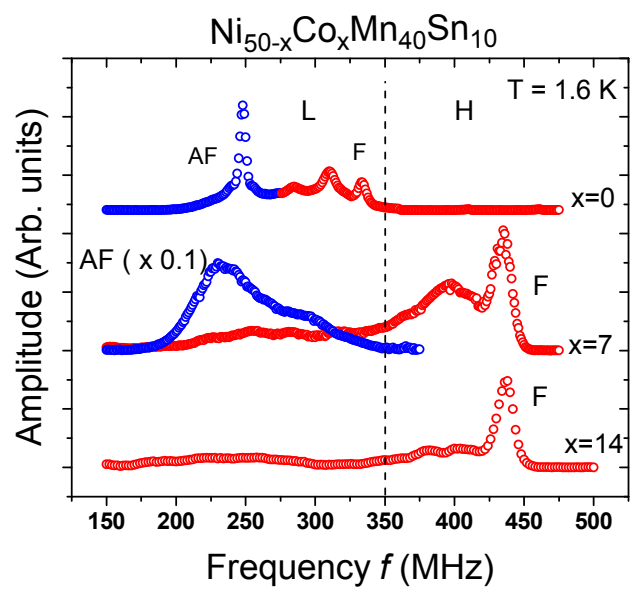

**FIG. 1**



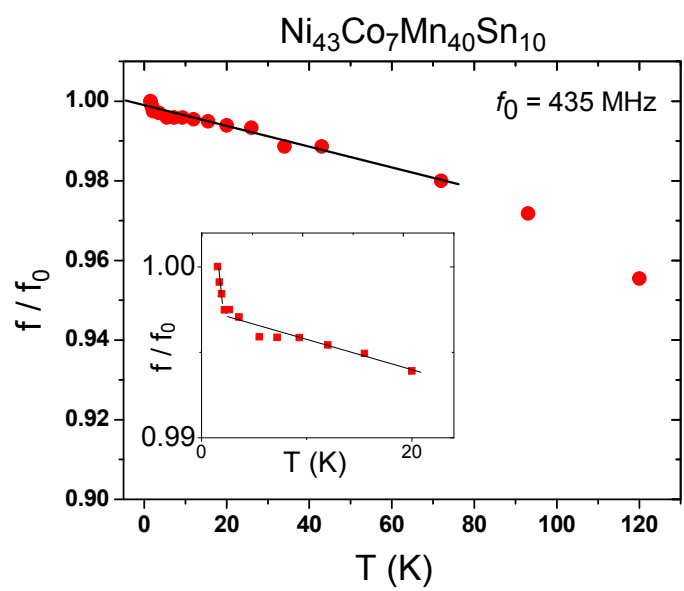

**FIG. 2**



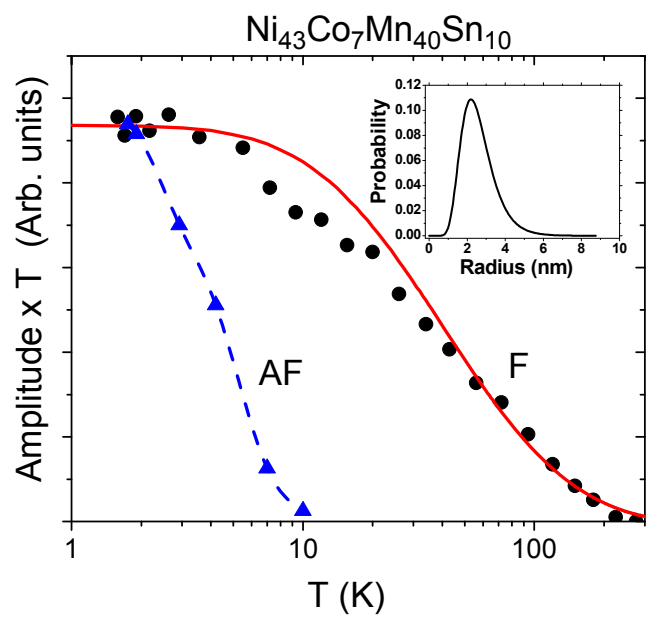





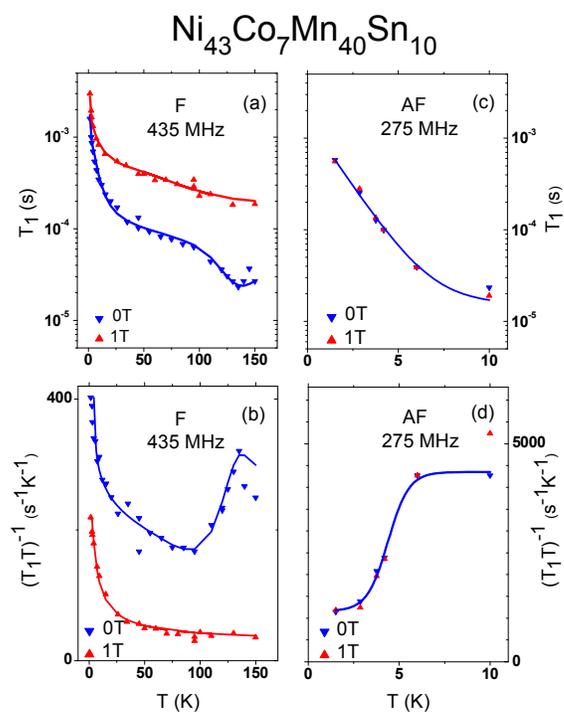





$Ni_{43}Co_7Mn_{40}Sn_{10}$

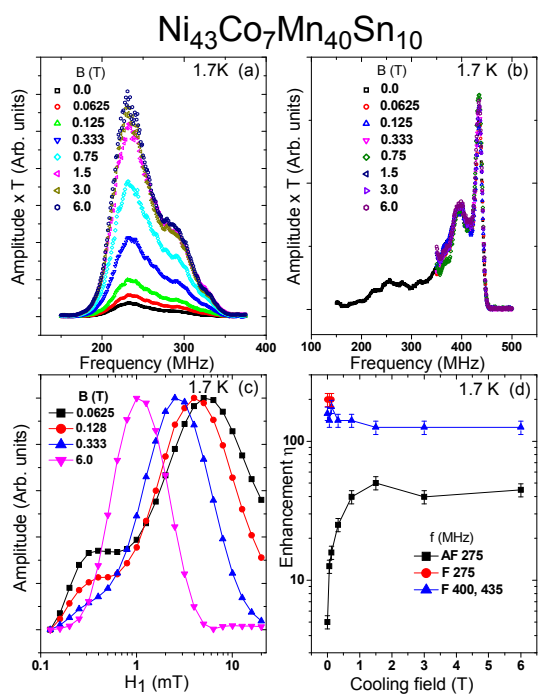

**FIG. 5**